\documentclass[preprint,aps,nofootinbib]{revtex4}
\usepackage{graphicx}% Include figure files
\usepackage{latexsym}
\usepackage{amssymb}
\usepackage{epsfig}
\usepackage{bm}% bold math
\usepackage{amsmath}
\newcommand{\be}{\begin{equation}}
\newcommand{\ee}{\end{equation}}
\newcommand{\beqq}{\setlength\arraycolsep{2pt}\begin{eqnarray}}
\newcommand{\eeqq}{\vspace{0cm} \end{eqnarray}}
\newcommand{\bea}{\begin{eqnarray}}
\newcommand{\eea}{\end{eqnarray}}

\newcommand{\bn}{\begin{eqnarray}}
\newcommand{\en}{\end{eqnarray}}

\newcommand{\nn}{\nonumber}

\newcommand{\no}{\noindent}

\def\bea{\begin{eqnarray}}
\def\eea{\end{eqnarray}}

\newcommand{\beq}{\begin{eqnarray}}
\newcommand{\eeq}{\end{eqnarray}}
\begin{document}

\title{New cosmological solutions in massive gravity theory}

\author{A. Pinho S. S.} \email{alexandre.pinho510@gmail.com}
\author{S. H. Pereira} \email{shpereira@feg.unesp.br}
\author{E. L. Mendon\c ca} \email{elias@gmail.com}

\affiliation{Faculdade de Engenharia de Guaratinguet\'a \\ UNESP - Univ. Estadual Paulista ``J\'ulio de Mesquita Filho''\\ Departamento de F\'isica e Qu\'imica\\ Av. Dr. Ariberto Pereira da Cunha 333 - Pedregulho\\
12516-410 -- Guaratinguet\'a, SP, Brazil}

%\maketitle

\pacs{95.35.+d, 95.36.+x, 98.80.$\pm$k, 12.60.$\pm$i}
\keywords{Dark matter, Dark energy, Cosmology, Models beyond the standard model}

%\bigskip
\begin{abstract}
In this paper we present some new cosmological solutions in massive gravity theory. Some homogeneous and isotropic solutions correctly describe accelerated evolutions for the universe. The study was realized considering a specific form to the fiducial metric and found different functions and constant parameters  of the theory that guarantees the conservation of the energy momentum tensor. Several accelerating cosmologies were found, all of them reproducing a cosmological constant term proportional to the graviton mass, with a de Sitter type solution for the scale factor.  We have also verified that when the fiducial metric is close to the physical metric the solutions are absent, except for some specific open cases.
\end{abstract}

\maketitle

%%%%%%%%%%%%%%%%%%%%%%%%%%%%%%%%%%%%%%%%%%%%%%%%%%%%%%%%%%%%%%%%%%%%%%%%%%

\section{Introduction}

The idea to give a non-null mass to graviton is an old one in the history of physics \cite{fierz,deser}. The proof of the existence of a non-linear generalization of the so called massive gravity theory is a problem that has stimulated several studies in last years \cite{hassan1,hassan2,hassan3,rham,hinterRMP,massiveG,volkov0,volkov,koba,gumru1,gumru2,rham2,dubowski, matas,ref05}, since the current observations of Supernovae type Ia (SNIa) \cite{SN,union2}, Cosmic
Microwave Background (CMB) radiation \cite{WMAP,planck} and Hubble parameter data \cite{farooq,sharov} indicate an accelerated expansion of the universe. Massive gravitons could perfectly mimic the effect of a cosmological constant term, rendering the theory a good alternative to the $\Lambda$CDM model of cosmology, which is plagued with several
fundamental issues related to the cosmological constant in a Friedmann-Lemaître-Robertson-Walker (FLRW) background \cite{CC}.

In general, massive gravity theories have ghost instabilities at the non-linear order, known as Boulware-Deser (BD) ghosts \cite{deser}. A procedure recently outlined in \cite{rham} successfully obtained actions that are ghost-free for the fully non-linear expansion, resulting in theories sometimes called dRGT (de Rham-Gabadadze-Tolley) model, that was first shown to be BD ghost free by Hassan et al. \cite{hassan1,hassan2,hassan3}. Massive gravity is constructed with an additional metric $f_{\mu\nu}$ (called fiducial metric) together the physical metric $g_{\mu\nu}$. When the fiducial metric is also endowed with a dynamic, the theory is called bimetric gravity or just bigravity. In such theory the problems of ghosts and instabilities are absent, but a new mass scale related to the fiducial metric is also present in addition to the graviton mass. For this reason we prefer to study the massive gravity version where just the graviton mass is present, and the reference metric is assumed to be of non-FLRW type, admitting non-isotropic and non-homogeneous forms. In some sense we study similar cases already presented in \cite{volkov0}, but several others solutions are also analysed, including an approach based just on the conservation of the energy momentum tensor. In some sense we have also generalized the cases of \cite{volkov0}, since that we have considered a physical metric of Friedmann type.

It is well known that when the fiducial metric is assumed to be flat, isotropic flat and closed FLRW cosmologies do not exist even at the background level. On the other hand isotropic open cosmologies exist as classical solutions but there are also perturbations that are unstable \cite{gumru1,gumru2}. Quantum cosmological models in the framework of nonlinear massive gravity has also been studied in \cite{vakili}, but some constraints present in the theory prevents it to admits flat and closed FLRW solution, although it is not the case for the open universe. Cosmological constraints with observational data for massive gravity theories have also been studied recently \cite{saridakis,nosso}, allowing the determination of free parameters of the theory. 

The paper is organized as follows. In Section II we present the general massive gravity theory. In Section III we present several cases with accelerating solutions based on the knowing of the fiducial metric. In Section IV we present solutions based on the conservation of the energy momentum tensor. We conclude in Section V.

\section{Massive gravity theory without ghosts}

The covariant action for massive gravity theory according to \cite{rham} can be written as:

\bea S=\frac{1}{8\pi G}\int d^4x\,\,\sqrt{-g}\left\lbrack -\frac{R}{2}+m^2{\cal U}({\cal K})\right\rbrack + S^{(mat)} \label{complete}\,,
\eea
with $S^{(mat)}$ representing the ordinary matter contend (radiation, baryons, dust, etc) and
\begin{equation}
{\cal U}({\cal K})={\cal U}_2+\alpha_3{\cal U}_3+\alpha_4{\cal U}_4
\label{U}
\end{equation}
acting like a potential due to the graviton mass term, $m$. In (\ref{U}) $\alpha_3$ and $\alpha_4$ are constant parameters and the functions ${\cal U}_2$, ${\cal U}_3$ and ${\cal U}_4$ are given by:
\begin{eqnarray}{\cal U}_2&=&\frac{1}{2!}([{\cal K}]^2-[{\cal K}^2])\\
{\cal U}_3&=&\frac{1}{3!}([{\cal K}]^3-3[{\cal K}][{\cal K}^2]+2[{\cal K}^3])\\
{\cal U}_4&=&\frac{1}{4!}([{\cal K}]^4-6[{\cal K}]^2[{\cal K}^2]+3[{\cal K}^2]^2+8[{\cal K}][{\cal K}^3]-6[{\cal K}^4])\,,
\end{eqnarray}
where traces are represented by $[{\cal K}]$ and superior orders of ${\cal K}^n (n=2,3,4)$ are ${{\cal K}^{\mu}}_{\alpha}{{\cal K}^{\alpha}}_{\nu}$, ${{\cal K}^{\mu}}_{\alpha}{{\cal K}^{\alpha}}_{\beta}{{\cal K}^{\beta}}_{\nu}$ and ${{\cal K}^{\mu}}_{\alpha}{{\cal K}^{\alpha}}_{\beta}{{\cal K}^{\beta}}_{\gamma}{{\cal K}^{\gamma}}_{\nu}$, respectively.   

We introduce ${\cal K}$ as
\bea
 {{\cal K}^{\mu}}_{\nu}={\delta ^{\mu}}_{\nu}-{\gamma ^{\mu }}_{\nu}
\label{K} 
\eea 
with $\gamma$ given by
\bea
{\gamma ^{\mu }}_{\nu}=\sqrt{{g}^{\mu\alpha}{f}_{\alpha\nu }},
\label{gammad}
\eea
where $g$ stands for the physical metric and $f$ represents the fiducial metric. The matrix represented by ${{\cal K}^{\mu}}_{\nu}$ and consequently the mass term breaks general covariance due to the presence of the fiducial metric. However thanks to the well-known St\"{u}ckelberg trick \cite{massiveG} one can think $f_{\mu\nu}$ as a covariant tensor field which can be constructed as follow:

\be f_{\mu\nu}=\eta_{AB}\partial_{\mu}X^A\partial_{\nu}X^B\ee

\no where $X^{A}$ are a set of four fields which transforms as scalars under general coordinate transformation of spacetime and are called St\"{u}ckelberg scalars and at the same time $\eta_{AB}=$diag$[1,\,-1,\,-1,\,-1]$. Notice that when the physical metric and the fiducial metric coincide we have ${{\cal K}^{\mu}}_{\nu}$ identically zero, which cancel the contribution of massive gravity. In order to have a non-trivial contribution from the graviton mass term we must consider a fiducial metric slightly different from the physical one.

The variation of the action with respect to $g^{\mu\nu}$ results in the Einstein equations
\begin{equation}
G^{\mu}_{\nu}=m^2 T^{\mu}_{\nu}+8\pi G {T^{(mat)}}^{\mu}_{\nu}\,,
\end{equation}
where ${T^{(mat)}}^{\mu}_{\nu}$ corresponds to the energy-momentum tensor of the ordinary matter,
\begin{equation}
{T^{(mat)}}^{\mu}_{\nu}={1\over 8\pi G}\textrm{diag}[\rho(t),\,-p(t),\,-p(t),\,-p(t)]\,,
\end{equation}
and $T^{\mu}_{\nu}$ represents the contribution to the energy-momentum tensor coming from the graviton mass term,
\begin{equation}
T_{\mu\nu}=2\frac{\delta{\cal U}}{\delta g^{\mu\nu}}-{\cal U}g_{\mu\nu}
\label{Tmunu}
\end{equation}
Substituting (\ref{U}) in (\ref{Tmunu}) one finds
\bea
{{T}^{\mu}}_{\nu}&=&{{\gamma}^{\mu}}_{\alpha}\left\{{{{\cal K}}^{\alpha}}_{ \nu}-\left[{\cal K}\right]{{\delta}^{\alpha}}_{\nu}\right\}-{\alpha}_{3}{{ \gamma}^{\mu}}_{\alpha}\left\{{{\cal U}}_{2}{{\delta}^{\alpha}}_{\nu}-\left[ {\cal K}\right]{{{\cal K}}^{\alpha}}_{\nu}+{{\left({{\cal K}}^{2}\right)}^{ \alpha}}_{\nu}\right\}\nn\\
&&-{\alpha}_{4}{{\gamma}^{\mu}}_{\alpha}\left\{{{\cal U} }_{3}{{\delta}^{\alpha}}_{\nu}-{{\cal U}}_{2}{{{\cal K}}^{\alpha}}_{\nu }-\left[{\cal K}\right]{{\left({{\cal K}}^{2}\right)}^{\alpha}}_{\nu}+{{\left({ {\cal K}}^{3}\right)}^{\alpha}}_{\nu}\right\}-{\cal U}{{\delta}^{\mu}}_{\nu}.
\label{T_e_m}\eea
By considering that there is no interaction between ordinary matter and the massive graviton we have that the energy-momentum tensor must be conserved separately,
\begin{equation}
\nabla^{\mu}T^{(mat)}_{\mu\nu}=0,\hspace{1cm}\nabla^{\mu}T_{\mu\nu}=0\,,\label{em_tensor}
\end{equation}
which will impose conditions on the St\"{u}ckelberg fields and also in the $\alpha$'s parameters. 

In which follows we will considerer just the graviton mass term  contribution to the energy momentum tensor, present in the fiducial metric. The matter part can be added in the standard way.

\section{St\"{u}ckelberg's Analysis: Fiducial Metric}
When we are looking for proper St\"{u}ckelberg fields the starting point is to define the physical and the fiducial metric. In order to reproduce cosmological results we define that physical metric must be homogeneous and isotropic, i.e.:

\bea
{ ds }_{ g }^{ 2 }={ N(t) }^{ 2 }{ dt }^{ 2 }-{ a(t) }^{ 2 }{ dr }^{ 2 }-{ a(t) }^{ 2 }{ f(r) }^{ 2 }\left( { d\theta  }^{ 2 }+\sin ^{ 2 }{ \theta  } { d\phi  }^{ 2 } \right) ,
\label{phy_metric}\eea 
where $N(t)$ is known as lapse function and $f(r)$ can be equal to $r$, $sinh(r)$ or $sin(r)$ depending on what kind of universe we are dealing with, flat ($k=0$), open ($k=-1$) or closed ($k=+1$), respectively. As in \cite{volkov0}, we assume a generalized form to the fiducial metric, spherically symmetric:

\bea
{ds}_{f}^{2}&=&\left({\dot{T}(t,r)}^{2}-{\dot{U}(t,r)}^{2}\right){dt}^{2 }+2\left(\dot{T}(t,r) {T}^{\prime}(t,r)-\dot{U}(t,r)U^{\prime}(t,r)\right) dtdr+\nn\\
&&+\left({{T}^{\prime}(t,r)}^{2}-{{U}^{\prime}(t,r)}^{2}\right){dr}^{2}-{U(t,r) }^{2}\left({d\theta}^{2}+\sin ^{2}{\theta}{d\phi}^{2}\right).
\label{fid_metric}\eea

Here the functions $T(t,r)$ and $U(t,r)$ are closely related to the St\"{u}ckelberg fields. Besides, dots and primes represent a derivatives with respect $t$ and $r$, respectively. The ghost-free massive gravity introduced in the last section can be rewritten in terms of such functions through function (\ref{gammad}). We bind both metrics (\ref{phy_metric}) and (\ref{fid_metric}) according to (\ref{gammad}) and then we are able to describe dRGT theory using functions $T(t,r)$ and $U(t,r)$. 

From now on we assume the condition
\bea
\dot{T}(t,r) {T}^{\prime}(t,r)=\dot{U}(t,r)U^{\prime}(t,r),
\label{diag}
\eea
which guarantees that fiducial metric (\ref{fid_metric}) is diagonal and consequently homogeneous and symmetric, depending on the functions $T(t,r)$ and $U(t,r)$. With such constraint one can check, following (\ref{gammad}), that:

\bea
{ \left( { { \gamma  }^{ \mu  } }_{ \nu  } \right)  }^{ 2 }=\begin{bmatrix} \frac { A(t,r) }{ { N(t) }^{ 2 } }  & 0 & 0 & 0 \\ 0 & \frac { B(t,r) }{ { a(t) }^{ 2 } }  & 0 & 0 \\ 0 & 0 & \frac { { U(t,r) }^{ 2 } }{ { a(t) }^{ 2 }{ f(r) }^{ 2 } }  & 0 \\ 0 & 0 & 0 & \frac { { U(t,r) }^{ 2 } }{ { a(t) }^{ 2 }{ f(r) }^{ 2 } } \label{gamma2} \end{bmatrix},
\eea
where we have defined $A(t,r)= {\dot{T}(t,r)}^{2}-{\dot{U}(t,r)}^{2}$ and $B(t,r)={ {U}^{\prime}(t,r)}^{2}-{{T}^{\prime}(t,r)}^{2}$. We assume that there is a square root of (\ref{gamma2}) given by the ansatz:

\bea
{ { \gamma  }^{ \mu  } }_{ \nu  }=\begin{bmatrix} \frac { \pm \sqrt { A(t,r) }  }{ { N(t) } }  & F(t,r) & 0 & 0 \\ G(t,r) & \frac { \pm \sqrt { B(t,r) }  }{ { a(t) } }  & 0 & 0 \\ 0 & 0 & \frac { { U(t,r) } }{ { a(t) }{ f(r) } }  & 0 \\ 0 & 0 & 0 & \frac { { U(t,r) } }{ { a(t) }{ f(r) } } \label{gamma} \end{bmatrix},
\eea
where $G(t,r)$ and $F(t,r)$ are a priori undefined functions. But by squaring (\ref{gamma}) we see that such functions cannot assume any form. Therefore we have three different cases to analyse. The first one concerns a diagonal fiducial metric ($F(r,t)=G(t,r)=0$) and other two are non-diagonal ($F(r,t)\neq0$ and $G(t,r)=0$ or $F(r,t)=0$ and $G(t,r)\neq0$).

In order to check the viability of such choices for the St\"{u}ckelberg fields we will consider the conservation of energy-momentum tensor, $\nabla_\mu {T^\mu}_\nu=0$, for each component. Due to the specific form of (\ref{gamma}) we have ${\nabla}_{\mu}{{T}^{\mu}}_{2}={\nabla}_{\mu}{{T}^{\mu}}_{3}=0$. The condition of conservation of the $T^\mu_0$ component will be the same for all the cases following, it is given by:

\scriptsize
\bea
{\nabla}_{\mu}{{T}^{\mu}}_{0}&=&\Big[ -\frac{3}{a(t)} \frac{\sqrt{A(t,r)}}{ N(t)} \left( 3+3{\alpha}_{3}+{\alpha}_{4} \right)-\frac{2}{ {a(t)}^{2}}\frac{\left(-2U(t,r)+\sqrt{B(t,r)} f(r) \right) \sqrt{A(t,r)}}{f(r) N(t)} (1+2{\alpha}_{3}+{\alpha}_{4})+\nn\\
&&+\frac{1}{{a(t)}^{3}}\frac{\left(-U(t,r)+2B(t,r)f(r)\right)\sqrt{A(t,r)} U(t,r)}{{f(r)}^{2} N(t)} ({\alpha}_{3}+{ \alpha}_{4})\Big]\frac{da(t)}{dt}+\nn\\
&&-\frac{1}{2 a(t)}\frac{\left(-4\left( \frac{\partial U(t,r)}{\partial t}\right)\sqrt{B(t,r)}+\left(\frac{\partial B(t,r)}{\partial t}  \right)f(r)\right)}{f(r) \sqrt{B(t,r)}} \left(3+3{\alpha}_{3}+{\alpha}_{4} \right)+\nn\\
&&+\frac{1}{{a(t)}^{2}}\frac{\left(-2U(t,r)\left(\frac{\partial U(t,r)}{\partial t}\right) \sqrt{B(t,r)} +U(t,r)\left(\frac{\partial B(t,r)}{\partial t}\right)f(r)+2B(t,r)\left(\frac{\partial U(t,r)}{\partial t}\right)f(r)\right)}{{f(r)}^{2}\sqrt{B(t,r)}}(1+2{\alpha}_{3}+{\alpha}_{4})+\nn\\
&&-\frac{1}{2{a(t)}^{3}}\frac{\left(U(t,r)\left(\frac{\partial B(t,r)}{\partial t}\right)+4B(t,r)\left( \frac{\partial U(t,r)}{\partial t}\right)\right)U(t,r)}{{f(r)}^{2}\sqrt{B(t,r)}}({ \alpha}_{3}+{\alpha}_{4}). \label{NablaT0} 
\eea
\normalsize

In which follows we present the analysis for the conservation of the energy momentum tensor for several cases, according to the values of $\alpha_{3}$ and $\alpha_{4}$ or $A(t,r)$, $B(t,r)$, $F(t,r)$ and $G(t,r)$, depending on which case we are analysing.

\subsection{Diagonal metric}\label{CASE1}

Here we consider that $\gamma$, given in (\ref{gamma}), is diagonal ($G(t,r) = F(r,t)=0$) and the functions $A(t,r)$ and $B(t,r)$ are independents. However we still have the constraint (\ref{diag}). There are several functions which respect such constraint, but we are interested in cases for which ${\nabla}_{\mu}{{T}^{\mu}}_{0}=0$ (given by (\ref{NablaT0})) and ${\nabla}_{\mu}{{T}^{\mu}}_{1}=0$, which is given by:

\scriptsize
\bea 
{\nabla}_{\mu}{{T}^{\mu}}_{1}&=&\frac{1}{2}\frac{\left(\frac{\partial A(t,r) }{\partial r}\right)}{N(t)\sqrt{A(t,r)}}\left(3+3{\alpha}_{3}+{ \alpha}_{ 4 }\right)+\frac{2}{a(t)}\frac{\left(\left(\frac{df(r)}{dr}\right)\sqrt{ B(t,r)}+\left(\frac{\partial U(t,r)}{\partial r}\right)\right)}{f(r)}\left(3{\alpha}_{ 2 }+3{\alpha}_{3}+{\alpha}_{4}\right)+\nn\\
&& -\frac{1}{a(t)}\frac{\left(2A(t,r)\left(\frac{\partial U(t,r)}{\partial r}\right)+2\left(\frac{df(r)}{dr}\right)\sqrt{B(t,r)} A(t,r)+U(t,r)\left(\frac{\partial A(t,r)}{\partial r}\right)\right)}{f(r)N(t)\sqrt{A(t,r) }}(1+2{\alpha}_{3}+{\alpha}_{4})+\nn\\
&& -\frac{2}{{a(t)}^{2}}\frac{ \left(\left(\frac{df(r)}{dr}\right)\sqrt{B(t,r)}+\left(\frac{\partial U(t,r)}{\partial r}\right)\right)}{{f(r)}^{2}}(1+2{\alpha}_{3}+{\alpha}_{4})+\nn\\
&&+\frac{1}{{2a(t)}^{2}}\frac{\left(4A(t,r)\left(\frac{\partial U(t,r)}{\partial r}\right) +U(t,r)\left(\frac{\partial A(t,r)}{\partial r}\right) +4\left(\frac{df(r)}{dr}\right) \sqrt{B(t,r)} A(t,r)\right)U(t,r)}{{f(r)}^{2}N(t)\sqrt{A(t,r)}}({\alpha}_{3}+{ \alpha}_{4}).\label{NablaT1a}
\eea
\normalsize

We analyse some different cases in which follows.

\subsubsection{$A(t,r)=0$, $B(t,r)=4/f(r)^6$ and $U(t,r)=1/f(r)^2$}\label{A1}

The conservation condition (\ref{NablaT0}) gives an expression for $T(t,r)$, which depends on what kind of universe we are dealing with. One can check that $T(t,r)$ is a constant for $f(r)=r$ (flat universe), $T(t,r)=\pm 2\coth(r)^{2}$ for $f(r)=\sinh(r)$ (open universe) and $T(t,r)=\pm 2i\cot(r)^{2} $ for $f(r)=\sin(r)$ (closed universe). Therefore, the equations of motion obtained by varying the action with respect to $a(t)$ and $N(t)$ are, respectively:
\bea
\frac{4\ddot{a}(t)}{a(t)}+\frac{2{\dot{a}}(t)^{2}}{{a(t)}^{2}}+\frac{2k}{ {a(t)}^{2}}+\frac{{m}^{2}}{{a(t)}^{2}{f(r)}^{6}}\left(1+{2\alpha}_{3}+{\alpha}_{4}\right)-{m}^{2}\left(6+{4\alpha}_{3}+{ \alpha}_{4}\right)=0
\label{A11}
\eea
and
\bea 
\frac{6{\dot{a}}(t)^{2}}{{a(t)}^{2}}&+&\frac{6k}{{a(t)}^{2}}+\frac{{3m}^{2 }}{{a(t)}^{2}{f(r)}^{6}}\left(1+{2\alpha}_{3}+{\alpha}_{4}\right)
-\frac{{2m}^{2}}{{a(t)}^{3}{f(r)}^{9}}\left({\alpha}_{3}+{\alpha}_{4} \right)\nn\\
&-&{m}^{2}\left(6+{4\alpha}_{3}+{\alpha}_{4}\right)=0.
\label{A12}
\eea
These are the FLRW equations obtained for the particular choice of $A(t,r),\,B(t,r)$ and $U(t,r)$. The presence of the anisotropic terms with $f(r)$ makes the solutions of little interest for cosmology. However if we choose $\alpha_3=-1$ and $\alpha_4=1$, the solutions turns to be isotropic with the presence of a cosmological constant like term, related to the massive graviton mass,
\bea 
\frac{2\ddot{a}(t)}{a(t)}+\frac{{\dot{a}}(t)^{2}}{{a(t)}^{2}}+\frac{k}{{a(t)}^{2}}-{3\over 2}m^2=0\label{A12a}
\eea
\bea
\frac{{\dot{a}}(t)^{2}}{{a(t)}^{2}}+\frac{k}{{a(t)}^{2}}-{{m}^{2}\over 2}=0.
\label{A12b}
\eea
Such solutions are accelerating and could perfectly reproduce cosmological observations.

\subsubsection{$U(t,r)=0$}\label{A3}
In this case we have $T(t,r)=constant$, which leads to $A(t,r)=B(t,r)=0$. The energy-momentum tensor is always conserved and both the fiducial metric and $\gamma$ are null. Although it seems a pathological case, it just reproduces a cosmology with a cosmological constant, represented by the $m^2$ term, which comes from ${\cal U}(t,r)$ in the last term of (\ref{T_e_m}). The equations of motion are: 
\bea
\frac{2\ddot{a}(t)}{a(t)}+\frac{{\dot{a}}(t)^{2}}{{a(t)}^{2}}+\frac{k}{ {a(t)}^{2}}-{{m}^{2}\over 2}(6+4\alpha_3+\alpha_4)=0\label{Au01}
\eea
and
\bea
\frac{{\dot{a}}(t)^{2}}{{a(t)}^{2}}+\frac{k}{{a(t)}^{2}}-{{m}^{2}\over 6}(6+4\alpha_3+\alpha_4)=0.\label{Au02}
\eea

\subsubsection{$A(t,r)=W(t)^2$, $B(t,r)=K(t)^2$ and $U(t,r)=-K(t)f(r)$}\label{A4}

When one tries to impose conservation of the energy-momentum tensor it is found that the universe must be open $(k=-1)$ and the relation between $W(t)$ and $K(t)$ is described by
\bea
K(t)=-\int^{t}{\frac{W(\tilde{t})}{N(\tilde{t})}\dot{a}(\tilde{t})d\tilde{t}}.
\label{K_t}
\eea
However, the relation (\ref{K_t}) together with the conditions for $T(t,r)$ results in $\dot{a}(t)=-N(t)$, which has no cosmological interest. As can be seen from equation (\ref{K_t}), when $W(t)=N(t)$, $K(t)$ would be equal to $-a(t)$ and this results in a standard Friedmann metric. This is due the fact that there is no cosmology when fiducial metric mimics the Friedmann one \cite{rham}. 

\subsubsection{$A(t,r)=\frac{N(t)^2}{a(t)^2}B(t,r)$}

With this condition we intend to verify the compatibility of cosmology and massive gravity for Friedmann like metrics. Such compatibility have been already checked in \cite{rham2} and states that only an open universe could afford both theories. In some sense these analysis is also present in \cite{volkov0}, but the scale factor $a(t)^2$ present in the time component of the physical metric used in \cite{volkov0} does not represents a true FLRW metric. Here we have addressed a similar analysis but we have taken $N(t)^2$ in (\ref{phy_metric}), and after doing calculations such function $N(t)$ is assumed equal to one in order to recover the FLRW background. 

Here we present another approach to analyse solution close to Friedmann metric. Our choice is to find solutions of $T(t,r)$ and $U(t,r)$ and then try to manipulate them in order to get conservative cases. Besides we also find results that conserve energy momentum tensor like previous subsection. By confronting both results we conclude that there is really something special in theory that exclude solutions for fiducial metric to be isotropic and homogeneous.

There are two constraints on the functions $T(t,r)$ and $U(t,r)$ which guarantee diagonality of $\gamma^2$. By making ${\left({{\gamma}^{0}}_{0 }\right)}^{2}={\left({{\gamma}^{1}}_{1}\right)}^{2}$ in equation (\ref{gamma2}) we get $A(t,r)=\frac{N(t)^2}{a(t)^2}B(t,r)$, which give:
\bea
{\dot{T}(t,r)}^{2}-{\dot{U}(t,r)}^{2}=-\frac{{N(t)}^{2}}{{a(t)}^{2}} \left( {{T}^{\prime}(t,r)}^{2}-{{U}^{\prime}(t,r)}^{2}\right).
\label{Cond1}\eea
After substituting (\ref{diag}) in equation (\ref{Cond1}) one can find
\bea
 \dot{U}(t,r)=\pm\frac{N(t)}{a(t)}{T}^{\prime}(t,r)\,,
\label{U.}\eea
and
\bea
 \dot{T}(t,r)=\pm\frac{N(t)}{a(t)}U^{\prime}(t,r)\,.
\label{T'}\eea
By deriving (\ref{U.}) with respect to $t$ and (\ref{T'}) with respect to $r$  one can combine the results to get:
\bea
\ddot{U}(t,r)=\frac{a(t)}{N(t)}\frac{d}{dt}\left(\frac{N(t)}{a(t)}\right)\dot{U}(t,r)+\frac{{N(t)}^{2}}{{a(t)}^{2}}U^{\prime\prime}(t,r).
\label{U..}\eea
It is possible to check that $T(t,r)$ has the same solutions of $U(t,r)$ since that function satisfies an equation equivalent to (\ref{U..}). We have chosen $U(t,r)=R(r)Z(t)$ in order to separate functions of $r$ and $t$ in (\ref{U..}). For this solutions ${\cal C}$ will stand for the constant of Fourier method. 

When ${\cal C}=0$ the solution is
\bea
U(t,r)=\left( {a}_{1} \int ^{t}{ \frac{N(\tilde{t})}{a(\tilde{t})}}d\tilde{t} +{a}_{2} \right) \left( {a}_{3}+{a}_{4}r \right),
\label{C=0}\eea
where $a_1$, $a_2$, $a_3$ and $a_4$ are constants. 

For a positive constant (${\cal C}>0$) one obtains
\bea
U(t,r)&=&\left[ {b}_{1}\sinh{\left(\sqrt{ {\cal C}  }\int^{t}{\frac{N(\tilde{t})}{a(\tilde{t} )}}d\tilde{t}\right)}+{b}_{2}\cosh{\left(\sqrt{ {\cal C} }\int^{t}{\frac{ N(\tilde{t})}{a(\tilde{t})}}d\tilde{t}\right)}  \right]\times\nn\\
&&\left[{b}_{3}\sinh{\left(\sqrt{ {\cal C}  }r\right)}+{b}_{4}\cosh{\left(\sqrt{ {\cal C} }r\right)} \right],
\label{C>0}\eea
where $b_1$, $b_2$, $b_3$ and $b_4$ are constants. 

For a negative constant (${\cal C}<0$)
\bea
U(t,r)&=&\left[ {c}_{1}\sin{\left(\sqrt{\left| {\cal C} \right| }\int^{t}{\frac{N(\tilde{t})}{a(\tilde{t})}}d\tilde{t}\right)}+{c}_{2}\cos{\left(\sqrt{\left| {\cal C} \right| }\int^{t}{\frac {N(\tilde{t})}{a(\tilde{t})}}d\tilde{t}\right)}  \right]\times\nn\\
&&\left[ {c}_{3}\sin{\left(\sqrt{\left| {\cal C} \right| }r\right)}+{c}_{4}\cos{\left(\sqrt{\left| {\cal C} \right| }r\right)} \right],
\label{C<0}\eea
with $c_1$, $c_2$, $c_3$ and $c_4$ constants. 

By checking equations (\ref{C=0}), (\ref{C>0}) and (\ref{C<0}) it is possible to conclude that both functions $U(t,r)$ and $T(t,r)$ can be modeled in many ways according to constant values. It is worth to pay attention on constant ${\cal C}$, which is directly related to the curvature $k$ of the Friedmann metric. That is, after adjusting the constants $a_i$, $b_i$ and $c_i$ ($i=1,\,2,\,3,\,4$) to be zero we can make $R(r)$ to mimic $f(r)$.

For instance, when $a_3=0$ in (\ref{C=0}) we have $R(r)=r$ ($a_4$ can be absorbed by the temporal part) which is in according to $f(r)=r$ for $k=0$. In equation (\ref{C>0}), which corresponds to ${\cal C}>0$, it is possible to obtain $f(r)$ by doing ${\cal C}=1$. That choice recover $f(r)=\sinh(r)$ that represents $k=-1$. Last but not least, equation (\ref{C<0}) corresponds to the closed universe ($k=+1$) when ${\cal C}=-1$ and $c_4=0$. For such case $R(r)$ and $f(r)$ are equal to $\sin(r)$.

Therefore one can recover Friedmann behavior in the $r$ dependent part which can match a quasi-isotropic metric. Such behavior is desirable once an isotropic and homogeneous metric has already been proved to be compatible only with an open universe. 

Some specific cases related to this are:

\subsubsection*{4.1 \;\;\; $\pm \sqrt{A(t,r)}=N(t)u_1^2$, $\pm \sqrt{B(t,r)}=a(t)u_1^2$, $U(t,r)=a(t)f(r)u_1$}\label{B2}

For this case we have that $u_1$ is an arbitrary constant. The motion equations are given by

\bea
\frac { 2\ddot { a } (t) }{ a(t) } +\frac { { \dot { a }  }(t)^{ 2 } }{ { a(t) }^{ 2 } } +\frac { k }{ { a(t) }^{ 2 } } -{{ m }^{ 2 }\over 2}(6+ 4 \alpha_{ 3 }+{ \alpha  }_{ 4 } )+\nn\\
+{4{ m }^{ 2 }\over 3}\left( 3+{ 3\alpha  }_{ 3 }+{ \alpha  }_{ 4 } \right) { u_1 }-{{ m }^{ 2 }}\left( 1+{ 2\alpha  }_{ 3 }+{ \alpha  }_{ 4 } \right) { u_1 }^{ 2 }+\nn\\
+{{ m }^{ 2 }\over 6}\left( { \alpha  }_{ 4 } \right) { u_1 }^{ 4 }=0
\label{41a}
\eea
and
\bea
\frac{{\dot{a}}(t)^{2}}{{a(t)}^{2}}+\frac{k}{{a(t)}^{2}} -{{ m }^{ 2 }\over 6}(6+ 4 \alpha_{ 3 }+{ \alpha  }_{ 4 } )+\nn\\
+{2{ m }^{ 2 }\over 3}\left( 3+{ 3\alpha  }_{ 3 }+{ \alpha  }_{ 4 } \right) { u_1 }-{{ m }^{ 2 }}\left( 1+{ 2\alpha  }_{ 3 }+{ \alpha  }_{ 4 } \right) { u_1 }^{ 2 }+\nn\\
+{2{ m }^{ 2 }\over 3}\left( { \alpha  }_{ 3 }+{ \alpha  }_{ 4 } \right) { u_1 }^{ 3 }-{{ m }^{ 2 }\over 6}\left( { \alpha  }_{ 4 } \right) { u_1 }^{ 4 }=0,
\label{41b}
\eea

\subsubsection*{4.2 \;\;\; $\pm \sqrt{A(t,r)}=N(t)u_2^2$, $\pm \sqrt{B(t,r)}=a(t)u_2^2$, $U(t,r)=-a(t)f(r)u_2$}\label{B3}

In this case $u_2$ is a constant given by

\bea
u_2=\pm \sqrt{\frac{3+3{\alpha}_{3}+{\alpha}_{4}}{{\alpha}_{3}+{\alpha}_{4}}}.
\label{u2}
\eea

Friedmann equations are
\bea
\frac { 2\ddot { a } (t) }{ a(t) } +\frac { { \dot { a }  }(t)^{ 2 } }{ { a(t) }^{ 2 } } +\frac { k }{ { a(t) }^{ 2 } } -{{ m }^{ 2 }\over 2}(6+ 4 \alpha_{ 3 }+{ \alpha  }_{ 4 } )+\nn\\
+{{ m }^{ 2 }\over 3}\left( 1+{ 2\alpha  }_{ 3 }+{ \alpha  }_{ 4 } \right) { u_2 }^{ 2 }+{{ m }^{ 2 }\over 6}\left( { \alpha  }_{ 4 } \right) { u_2 }^{ 4 }=0
\label{B21}
\eea
and
\bea
\frac{{\dot{a}}(t)^{2}}{{a(t)}^{2}}+\frac{k}{{a(t)}^{2}}-{{m}^{2}\over 6}\left(6+{4\alpha}_{3}+{\alpha}_{4}\right)+\nn\\ 
+{{m}^{2}\over 3}\left(1+{2\alpha}_{3}+{\alpha}_{4}\right){u_2}^{2}-{{m}^{2 }\over 6}\left({\alpha}_{4}\right){u_2}^{4}=0,
\label{B22}
\eea

\subsubsection*{4.3 \;\;\; $\pm \sqrt{A(t,r)}=N(t)u_3^2$, $\pm \sqrt{B(t,r)}=-a(t)u_3^2$, $U(t,r)=-a(t)f(r)u_3$}\label{B4}

This case is more interesting due its resemblance with open universe case that allows compatibility with massive gravity. In such case \cite{gumru2} the fiducial metric is defined like $f_{\mu\nu}=\eta_{AB}\partial_{\mu}X^A\partial_{\nu}X^B$ with $X=(f(t)\sqrt{1+|K|(x^2+y^2+z^2)}$, $\sqrt{|K|}f(t)x$, $\sqrt{|K|}f(t)y$, $\sqrt{|K|}f(t)z)$. For such definition we have that $f(t)=\frac{a(t)}{\sqrt{|K|}}u_3$, where $u_3$ has the same value of $u$ from equation (\ref{u3}). Then we get the following pair of equations of motion:
\bea
\frac { 12\ddot { a } (t) }{ a(t) } +\frac { 6{ \dot { a }  }(t)^{ 2 } }{ { a(t) }^{ 2 } } +\frac { 6k }{ { a(t) }^{ 2 } } -3{ m }^{ 2 }\left( 6+{ 4\alpha  }_{ 3 }+{ \alpha  }_{ 4 } \right)+\nn\\
+2{ m }^{ 2 }\left( 1+{ 2\alpha  }_{ 3 }+{ \alpha  }_{ 4 } \right) { u_3 }^{ 2 }+{ m }^{ 2 }\left( { \alpha  }_{ 4 } \right) { u_3 }^{ 4 }=0
\label{B23}
\eea
and
\bea
\frac{6{\dot{a}}(t)^{2}}{{a(t)}^{2}}+\frac{6k}{{a(t)}^{2}}-{m}^{2}\left(6+{4\alpha}_{3}+{\alpha}_{4}\right)+\nn\\ +2{m}^{2}\left(1+{2\alpha}_{3}+{\alpha}_{4}\right){u_3}^{2}-{m}^{2 }\left({\alpha}_{4}\right){u_3}^{4}=0.
\label{B24}
\eea

\vspace{0.5cm}

In the next two sections we analyse massive gravity by considering off-diagonal elements in the function $\gamma$ given by (\ref{gamma}). In Section B it is considered that $G(t,r)$ is zero while $F(t,r)$ is non zero. Section D presents the opposite situation ($G(t,r)\neq 0$ and $F(t,r)=0$). For both cases it is imposed that $\sqrt{A(t,r)}=-\frac{N(t)}{a(t)}\sqrt{B(t,r)}$ in order to have ${\gamma}^{2}$ from (\ref{gamma2}) diagonal.

\subsection{Non-diagonal metric: Case 1}\label{CASE3}

As described previously, this case consists of making $F(t,r) \neq 0$ and $G(t,r)=0$. After substituting the condition for keeping ${\gamma}^{2}$ (\ref{gamma2}) diagonal into equation (\ref{NablaT0}) we get the condition to have ${\nabla}_{\mu}{{T}^{\mu}}_{0}$ null. The other term necessary to ensure conservation is given by the conservation of $T^\mu_1$, given by:

\scriptsize
\bea {\nabla}_{\mu}{{T}^{\mu}}_{1}&=&\left[-\frac{2F(t,r)}{a(t)}\left(3{\alpha}_{ 2}+3{\alpha}_{3}+{\alpha}_{4}\right)+\frac{2F(t,r)}{{a(t)}^{2}}\frac{U(t,r)}{ f(r)}(1+2{\alpha}_{3}+{\alpha}_{4})\right] \frac{da(t)}{dt}+\nn\\
&&-\frac{1}{2a(t)N(t)\sqrt{B(t,r)}}\Big[2\left(\frac{dN(t)}{dt}\right)F(t,r)a(t)\sqrt{B(t,r)}-\left(\frac {\partial B(t,r)}{ \partial r}\right)N(t)+\nn\\
&&+2\left(\frac{\partial F(t,r)}{\partial t}\right) N(t)a(t)\sqrt{B(t,r)}\Big]\left(3+3{\alpha}_{3 }+{\alpha}_{4}\right)+\nn\\
&&+\frac{1}{{a(t)}^{2}f(r)N(t)\sqrt{B(t,r)}}\Big[-2{\left(\frac{df(r)}{dr}\right) B(t,r)}^{{3}/{2}}N(t)+2\left(\frac{dN(t)}{dt}\right)F(t,r)U(t,r)a(t)\sqrt{B(t,r)} +\nn\\
&&+2F(t,r)\left(\frac{\partial U(t,r)}{\partial t}\right)a(t)N(t)\sqrt{B(t,r)}+2\left(\frac{\partial F(t,r)}{\partial t}\right)U(t,r)a(t)N(t)\sqrt{B(t,r)}-2B(t,r)\left( \frac{\partial U(t,r)}{\partial t}\right)N(t)+\nn\\
&&-U(t,r)\left(\frac{\partial B(t,r)}{\partial r }\right)N(t)\Big]\left(1+{2\alpha}_{3}+{ \alpha}_{4}\right) +\frac{2}{a(t)}\frac{\left(\left(\frac{\partial U(t,r)}{\partial r}\right)+\left(\frac{df(r)}{dr}\right)\sqrt{B(t,r)}\right)}{f(r)}\left(3{\alpha }_{2}+3{\alpha}_{3}+{\alpha}_{4}\right)+\nn\\
&&-\frac{1}{{a(t)}^{2}}\frac{\left(\left( \frac{\partial U(t,r)}{\partial r}\right)+\left(\frac{df(r)}{dr}\right)\sqrt{B(t,r)}  \right)U(t,r)}{{f(r)}^{2}}\left(1+{2\alpha}_{3}+{\alpha}_{4}\right)+\nn\\
&&+\frac{1}{{2a(t)}^{3}{f(r)}^{2}N(t)\sqrt{B(t,r)}}\Big[4\left(\frac{df(r)}{dr}\right){B(t,r)}^{{3 }/{2}}N(t)+4B(t,r)\left(\frac{\partial U(t,r)}{\partial r}\right)N(t)+\nn\\
&&-2\left(\frac{\partial F(t,r)}{\partial t}\right)U(t,r)a(t)N(t)\sqrt{B(t,r)}+U(t,r)\left(\frac{\partial B(t,r)}{\partial r}\right)N(t)+\nn\\
&&-2\left(\frac{dN(t)}{dt}\right) F(t,r)U(t,r)a(t)\sqrt{ B(t,r)}-4F(t,r)\left(\frac{\partial U(t,r)}{\partial t}\right)a(t)N(t)\sqrt{B(t,r)}  U(t,r)\Big]\left({\alpha}_{3}+{\alpha}_{4} \right).\label{NablaT1b}
\eea
\normalsize

Two different cases are analysed in which follow. 

\subsubsection{$A(t,r)=B(t,r)=0, F(t,r)=1/U(t,r)^2$, $G(t,r)=0$, $\alpha_3=-2$ and $\alpha_4=3$}\label{C1} 
With this choices there are some restrictions on functions $T(t,r)$ and $U(t,r)$. By having $A(t,r)$ and $B(t,r)$ null, the only condition is that both are constants or equal. So, it is possible to make $U(t,r)$ equal to $a(t)f(r)$, which would turn this metric closer to Friedmann. Anyway, equations of motion are given by
\bea
\frac{2\ddot{a}(t)}{a(t)}+\frac{{\dot{a}}(t)^{2}}{{a(t)}^{2}}+\frac{k}{ {a(t)}^{2}}-\frac{{m}^{2}}{2}=0
\label{eq1}
\eea
and
\bea
\frac{{\dot{a}}(t)^{2}}{{a(t)}^{2}}+\frac{k}{{a(t)}^{2}}-{{m}^{2}\over 6}=0.
\label{eq2}
\eea
Once again the massive gravity contribution comes as a cosmological constant in Friedmann equations. 
 
\subsubsection{$A(t,r)=u^2N(t)^2$, $B(t,r)=u^2a(t)^2$, $U(t,r)=-ua(t)f(r)$, $G(t,r)=0$ and $F(t,r)=ua(t)$}\label{C3}

Here $u$ is a constant to be determined. After some calculation we have the equations of motion:
\bea
\frac{12\ddot{a}(t)}{a(t)}+\frac{6{\dot{a}}(t)^{2}}{{a(t)}^{2}}+\frac{6k}{ {a(t)}^{2}}-3{m}^{2}\left(6+{4\alpha}_{3}+{\alpha}_{4}\right)+5{ m}^{2}u\left(3+{3\alpha}_{3}+{\alpha}_{4}\right)+\nn\\
-{m}^{2}{u}^{2}\left(1+{2\alpha}_{3}+{\alpha}_{4}\right)-{m}^{2}{u}^{3}\left({ \alpha}_{3}+{\alpha}_{4}\right)=0
\label{C31}
\eea
and
\bea
\frac{6{\dot{a}}(t)^{2}}{{a(t)}^{2}}+\frac{6k}{{a(t)}^{2}}-{m}^{2}\left( 6+{4\alpha}_{3}+{\alpha}_{4}\right)+{m}^{2}u\left(3+{ 3\alpha}_{3}+{\alpha}_{4}\right)+\nn\\
+{m}^{2}{u}^{2}\left(1+{2\alpha}_{ 3}+{\alpha}_{4}\right)-{m}^{2}{u}^{3}\left({\alpha}_{3}+{\alpha}_{4}\right)=0,
\label{C32}
\eea
where $u$ is given by
\bea
u=\frac { -4\left( 1+{ 2\alpha  }_{ 3 }+{ \alpha  }_{ 4 } \right) \pm \sqrt { 4{ \left( 1+{ 2\alpha  }_{ 3 }+{ \alpha  }_{ 4 } \right)  }^{ 2 }-4\left( { \alpha  }_{ 3 }+{ \alpha  }_{ 4 } \right) \left( { 3+3\alpha  }_{ 3 }+{ \alpha  }_{ 4 } \right)  }  }{ 2\left( { \alpha  }_{ 3 }+{ \alpha  }_{ 4 } \right)  } .\label{u3}
\eea  
With a particular choice of $\alpha_3=0$ and $\alpha_4=-3$ we have $u=-2$ or $u=+2/3$. Both cases furnish a positive cosmological constant to the mass term, which can also lead to an accelerated expansion.

\subsection{Non-diagonal metric: Case 2}\label{CASE4}

Here we follow the same idea of last subsection but we have $F(t,r)= 0$ and $G(t,r)\neq 0$. Therefore, anisotropy is still kept. Conditions for conservation are obtained after plugging $\sqrt{{A(t,r)}}=-\frac{{ N(t)}}{{a(t)}}\sqrt{B(t,r)}$ into (\ref{NablaT0}) and for the $T^\mu_1$ we have:

\scriptsize
\bea {\nabla}_{\mu}{{T}^{\mu}}_{1}&=&\Big[-\frac{a(t)G(t,r)}{{N(t)}^{2}}\left(3+3{\alpha}_{3}+{\alpha}_{4}\right)+\frac{2G(t,r)}{{N(t)}^{2}}\frac{ U(t,r)}{f(r)}(1+2{\alpha}_{3}+{\alpha}_{4})+\nn\\
&&+\frac{1}{a(t)}\frac{ 2G(t,r)}{{N(t)}^{2}}\frac{{U(t,r)}^{2}}{{f(r)}^{2}}({\alpha}_{3}+{\alpha}_{ 4})\Big]\frac{da(t)}{dt}+\frac{1}{2a(t)}\frac{\left(\frac{\partial B(t,r)}{ \partial r}\right)}{\sqrt{B(t,r)}}\left(3+3{\alpha}_{3}+{\alpha}_{ 4}\right)+\nn\\
&&+\frac{2}{a(t)}\frac{\left(\left(\frac{df(t,r)}{dr}\right)\sqrt{B(t,r)} +\left(\frac{\partial U(t,r)}{\partial r}\right)\right)}{f(r)}\left(3+3{ \alpha}_{3}+{\alpha}_{4}\right)+\nn\\
&&-\frac{1}{{a(t)}^{2}}\frac{\left( 2B(t,r)\left( \frac{\partial U(t,r)}{\partial r}\right)+2\left(\frac{df(t)}{dr}\right){B(t,r)}^{{3 }/{2}}+U(t,r)\left(\frac{\partial B(t,r)}{\partial r}\right)\right)}{f(r)\sqrt{B(t,r)}  }\left(1+2{\alpha}_{3}+{\alpha}_{4}\right)+\nn\\
&&-\frac{2}{{a(t)}^{2}} \frac{\left(\left(\frac{df(t)}{dr}\right)\sqrt{B(t,r)}+\left(\frac{\partial U(t,r)}{ \partial r}\right)\right)U(t,r)}{{f(r)}^{2}}\left(1+2{\alpha}_{3}+{ \alpha}_{4}\right)+\nn\\
&&+\frac{1}{{a(t)}^{3}}\frac{\left(4B(t,r)\left(\frac{\partial U(t,r)}{\partial r}\right)+U(t,r)\left(\frac{\partial B(t,r)}{\partial r}\right)+4\left( \frac{df(t)}{dr}\right){B(t,r)}^{{3}/{2}}\right)U(t,r)}{{f(r)}^{2}\sqrt{B(t,r) }}\left({\alpha}_{3}+{\alpha}_{4}\right). \label{NablaT1c}
\eea

\normalsize

We analyse just one case:

\subsubsection{$A(t,r)=B(t,r)=U(t,r)=F(t,r)=0$ and $\alpha_4=-3(\alpha_3+1)$}\label{D2}

This case is dictated by the fact that function (\ref{gamma}) has only one non zero element, that is off-diagonal. After squaring it, we see that equation (\ref{gamma2}) is identically zero. This means that ${\cal K}$ becomes an identity matrix in Lagrangian. $G(t,r)$ is an arbitrary function with $\alpha_4=-3(\alpha_3+1)$ being the only and just necessary condition for conservation of energy momentum tensor. From this we get the equations of motion,
\bea
\frac{2\ddot{a}(t)}{a(t)}+\frac{{\dot{a}}(t)^{2}}{{a(t)}^{2}}+\frac{k}{ {a(t)}^{2}}-{1\over 2}(3+{\alpha}_{3}){m}^{2}=0
\label{D21}
\eea
and
\bea
\frac{{\dot{a}}(t)^{2}}{{a(t)}^{2}}+\frac{k}{{a(t)}^{2}}-{1\over 6}(3+{\alpha}_{3}){m}^{2}=0.
\label{D22}
\eea
Both equations (\ref{D21}) and (\ref{D22}) show us that massive gravity makes a cosmological constant to appear in the Friedmann equations. For this case it is possible to eliminate gravitation contribution by making $\alpha_3=-3$.

\section{Conclusion}

In this paper we have obtained new cosmological solutions in massive gravity theory constructed by means of to set specific forms to the fiducial metric (\ref{fid_metric}) and found different functions $A(t,r),\,B(t,r),\,U(t,r)$ and $T(t,r)$ that guarantees the conservation of the energy momentum tensor. The equations for energy momentum tensor conservation were presented in the general form for three different cases, one being diagonal in the fiducial metric and two being non-diagonal. The expressions for the energy momentum conservation are (\ref{NablaT0}), (\ref{NablaT1a}), (\ref{NablaT1b}) and (\ref{NablaT1c}), being the equation for the $T^\mu_0$ component valid for all the cases treated here. In some cases the parameters $\alpha$'s were also specified in order to give accelerated solutions. A particular choice for the $\alpha$'s parameters also permits to turns an anisotropic solution into an isotropic one, as occurs in the first case, Eqs. (\ref{A11}-\ref{A12}), where the choices $\alpha_3=-1$ and $\alpha_4=1$ leads to the isotropic equations (\ref{A12a}-\ref{A12b}). We are interested in homogeneous and isotropic solutions in order to respect the cosmological principle. Anisotropic solution may be interesting in another specific cases, but for the current evolution of the universe we look forward to homogeneous and isotropic cases.

Several homogeneous and isotropic solutions that correctly describes accelerated evolutions for the universe were also found, all of them reproducing a cosmological constant term proportional to the graviton mass. From all the pair of equations (\ref{A12a}-\ref{A12b}), (\ref{Au01}-\ref{Au02}), (\ref{41a}-\ref{41b}), (\ref{B21}-\ref{B22}), (\ref{B23}-\ref{B24}), (\ref{eq1}-\ref{eq2}), (\ref{C31}-\ref{C32}) and (\ref{D21}-\ref{D22}) it is easy to see that all they can be reduced to an single equation of the form:
\bea
{\ddot{a}\over a}=C(\alpha,u)m^2\,,
\eea
where $c(\alpha,u)$ is a constant function depending on the $\alpha$'s free parameters and constant $u$'s. It is easy to see that the solution to the above equation is
\bea
a(t) = a_0 \exp(\pm \sqrt{C(\alpha,u)}mt)\,,
\eea  
where the positive solution is a de Sitter type solution, with the property to give an accelerated expansion to the universe, here sourced by the graviton mass $m$.

The constant free parameters $\alpha_3$ and $\alpha_4$ are the contributions coming from the lagrangians counter terms (\ref{U}) needed to eliminate Bouware-Deser ghost present in theory. From a classical point of view, the constants $\alpha$'s are yet unknown, but may be its values comes from a perturbative quantum construction of the theory. Although a theoretical approach in order to obtain the values of such constants are yet not available, observational constraints to its can be done \cite{nosso}.

We have also verified that when the fiducial metric is close to the physical metric the solutions are absent. As already shown for different works this is a characteristic of the massive gravity theory, which forbids the fiducial metric to be of the same type of the physical one.

%%%%%%%%%%%%%%%%%%%%%%%%%%%%%%%%%%%%%%%%%%%%%%%%%%%%%
\begin{acknowledgments}
The authors are grateful to referee
for several helpful comments concerning the paper. APSS thanks CAPES - Coordena\c{c}\~{a}o de Aperfei\c{c}oamento de Pessoal de N\'{i}vel Superior, for financial support. SHP is grateful to CNPq - Conselho Nacional de Desenvolvimento Cient\'ifico e Tecnol\'ogico, Brazilian research agency, for financial support, grants number 304297/2015-1. ELM is supported by CNPq under the grant (449806/2014-6).
\end{acknowledgments}
%%%%%%%%%%%%%%%%%%%%%%%%%%%%%%%%%%%%%%%%%%%%%%%%%%%%%

%\newpage
%%%%%%%%%%%%%%%%%%%%%%%%%%%%%%%%%%%%%%%%%%%%%%%%%%%%%%%%%%%%%%%%%%%%%%%%%%

\end{document}